\documentclass[prl,twocolumn,english,showpacs,superscriptaddress]{revtex4}
\usepackage{amsmath,amssymb,amsfonts}
\usepackage{CJK}
\usepackage{bm}
\usepackage{tikz}
\usepackage{pgfplots}
\usetikzlibrary{matrix,fit}
\usepackage{graphicx} 
\usepackage{braket}
\usepackage{subfig}
\usepackage{caption}
\usepackage{mathrsfs}
\usepackage{float}
\usepackage[utf8]{inputenc}
\usepackage{pgflibraryarrows}
\usepackage{lipsum} 
\usepackage{pgflibrarysnakes}

\begin{document}

\title{Second-Order Photonic Topological Insulator with Corner States}

\author{Bi-Ye Xie}
\affiliation{National Laboratory of Solid State Microstructures and Department of Materials Science and Engineering, Nanjing University, Nanjing 210093, China}
\author{Hong-Fei Wang}
\affiliation{National Laboratory of Solid State Microstructures and Department of Materials Science and Engineering, Nanjing University, Nanjing 210093, China}
\author{Hai-Xiao Wang}
\affiliation{School of Physical Science and Technology, and
  Collaborative Innovation Center of Suzhou Nano Science and
  Technology, Soochow University, 1 Shizi Street, Suzhou 215006,
  China}
\author{Xue-Yi Zhu}
\affiliation{National Laboratory of Solid State Microstructures and Department of Materials Science and Engineering, Nanjing University, Nanjing 210093, China}
\author{Jian-Hua Jiang}
\email[]{joejhjiang@hotmail.com}
\affiliation{School of Physical Science and Technology, and
  Collaborative Innovation Center of Suzhou Nano Science and
  Technology, Soochow University, 1 Shizi Street, Suzhou 215006,
  China}
\author{Ming-Hui Lu}
\email[]{luminghui@nju.edu.cn}

\affiliation{National Laboratory of Solid State Microstructures and Department of Materials Science and Engineering, Nanjing University, Nanjing 210093, China}
\affiliation{ Jiangsu Key Lab. of Artificial Functional Materials, Nanjing University, Nanjing 210093, China}
\affiliation{Collaborative Innovation Center of Advanced Microstructures, Nanjing University, Nanjing 210093, China}
\author{Yan-Feng Chen}
\affiliation{National Laboratory of Solid State Microstructures and Department of Materials Science and Engineering, Nanjing University, Nanjing 210093, China}


\begin{abstract}
Higher-order topological insulators (HOTIs) which go beyond the description of conventional bulk-boundary correspondence, broaden the understanding of topological insulating phases. Being mainly focused on electronic materials, HOTIs have not been found in photonic crystals yet. Here, we propose a type of two-dimensional second-order photonic crystals with zero-dimensional corner states and one-dimensional boundary states for optical frequencies. All of these states are topologically nontrivial and can be understood based on the theory of topological polarization. Moreover, by tuning the easily-fabricated structure of the photonic crystals, different topological phases can be realized straightforwardly. Our study can be generalized to higher dimensions and provides a platform for higher-order photonic topological insulators and semimetals.
\end{abstract}
\maketitle

\begin{center}
\textbf{\uppercase\expandafter{\romannumeral1}. INTRODUCTION}
\end{center}

Topological insulators (TIs) and topological semimetals (TSMs) have been theoretically and experimentally studied due to their distinct edge states and transport properties~\cite{TI1,TI2}. Normally, $d$-dimensional ($d$D) TIs  have $d$D gapped bulk states and $(d-1)$D gapless boundary states. Recently, the concept of higher-order topological insulator (HOTI) has been put forward to describe those topological insulators (TIs) which have lower-dimensional gapless boundary states~\cite{HOTI1,HOTI2,HOTI3,HOTI4,HOTI5,HOTI6}. Generally speaking, a $d$D TI with $(d-1)$D, $(d-2)$D, ..., $(d-n-1)$D gapped boundary states and $(d-n)$D gapless boundary states is called the $n$th-order TI. The HOTIs broaden the family of nontrivial topological insulating phases. Moreover, the HOTIs have unique boundary states which go beyond the conventional bulk-boundary correspondence and are characterized by novel topological invariants~\cite{HOTI1,HOTI2,HOTI4,HOTI6}. 

However, it is not easy to realize these topological phases in electronic materials. One of the obstacles is that the Fermi levels of electronic materials are not always in the topologically nontrivial band gaps or at the gapless points. The band structures of photonic crystals (PCs) provide us with platforms to study various topological phases such as photonic topological insulators (PTIs) and photonic topological semimetals (PTSMs) without limitations imposed by the Fermi level~\cite{PTI1,PTI2,PTI3,PTI4,PTI5,PTI6,PTI7,PTI8,PTI9,PTI10,PTI11,PTI12,PTI13,PTI14,PTI15,PTI16,PTI17,PTI18,PTI19,PTI20,PTSM1,PTSM2,PTSM3,PTSM4,PTSM5,PTSM6}. In terms of the HOTIs, the topological corner and hinge states in PCs can be used to design robust optical cavities and waveguides. So far, the observations of HOTIs are only realized in mechanical metamaterials~\cite{HOTIEXP1}, electrical circuits~\cite{HOTIEXP2,HOTIEXP3}, and weakly coupled optical waveguides~\cite{HOTIEXP4} which are described predominantly with quadrupole or rotation-symmetry-protected topological orders. The extension of the notion of HOTIs to PCs without negative coupling is still lacking.

In this paper, we propose a two-dimensional (2D) PTI which is the 2D photonic generalization of the Su-Schrieffer-Heeger
(SSH) model~\cite{SSH}. Similar to the 1D SSH model, the topological classes of the 2D photonic SSH model can be determined by different lattice structures as proposed in Ref. ~\cite{2DSSH0}. Previous studies of the 2D photonic SSH model is focused on the zero Berry curvature of the bulk band 
topology and gapless 1D edge states~\cite{2DSSH0,2DSSH1}. Here, by theoretical investigation and numerical simulation, we demonstrate that there are coexisting edge and corner states when two topologically distinct PCs are placed together to form box-shaped boundaries. We reveal that both the bulk polarization, described by the vector ${\bm P}=(P_x, P_y)$, and the edge polarization, $p_x^{\nu_y}$ and $p_y^{\nu_x}$, are  quantized by the mirror symmetries $M_{x}:=x\to -x$ and $M_y:=y\to -y$. The theoretical predictions and analysis are supported by numerical simulations for all-dielectric PCs at optical frequencies. 

This paper is organized as follows. The band structures of PC and topological corner states are introduced in Sec. II. In Sec. III, we study the 1D edge states and extend our discussion from isotropic case to anisotropic case where lattice constant along $x-$direction can be different from the one along $y-$direction. Finally, a summary is given in Sec. IV.

\begin{center}
\textbf{\uppercase\expandafter{\romannumeral2}. SECOND-ORDER TOPOLOGICAL PHOTONIC CRYSTALS}
\end{center}

\begin{figure}
\centering
\includegraphics[scale=0.21]{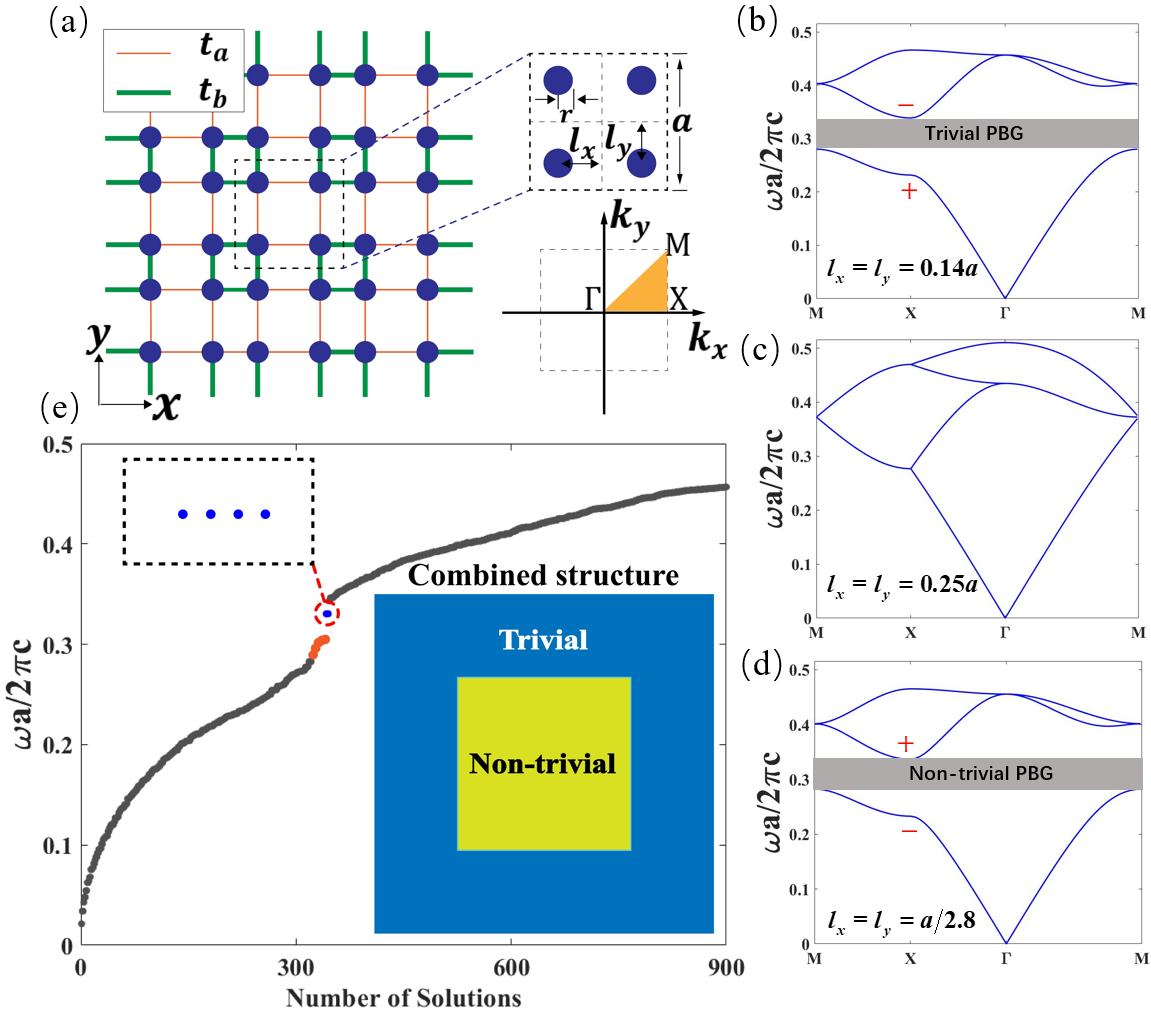}
\captionsetup{format=plain, justification=raggedright }
\caption{(a) The 2D photonic SSH model and the first Brillouin zone. We set $a=1.5\mu m$, $r=0.12a$ and the relative dielectric constant $\epsilon=12$ for all the cases. The coupling strength denoted as $t_a$ and $t_b$. (b)-(d) The band inversion induced by changing $l$. $+$ ($-$) indicates the even (odd) parity of the band. (b) $l=0.14a$ (topologically trivial PBG). (c) $l=0.25a$ (band gap closing). (d) $l=a/2.8$ (topologically nontrivial PBG). We choose the values of $l$ in order to make maximal overlap between the band gaps in (b) and (d). (e) Photonic eigenmodes of a combined structure where the PC in (d) is inside a box and the PC in (b) is outside the box, as shown in the inset. There are four degenerate states localized at the four corners in the band gap. Here edge (bulk) modes are denoted by orange (gray) points, whereas the corner modes are represented by blue points.}
\label{fig:1}
\end{figure}

We consider a 2D PC with mirror symmetries as shown in Fig.~\ref{fig:1}(a). There are four identical dielectric rods in each unit-cell which form isotropic ($l_x=l_y$) or anisotropic (with $l_x\neq l_y$) PCs, depending on their configurations (square or rectangular). The center of the configuration is at the center of the unit-cell which is also the origin of the coordinates throughout this paper. Since there is a reflection symmetry in $z-$direction, the electro-magnetic fields which are the eigenmodes of the PC can be classified into transverse-electric (TE) fields and transverse-magnetic (TM) fields. Without loss of generality, we consider the TM modes throughout our paper and the TE modes can be studied in a similar way. 

Photonic band structures of distinct topological properties can be realized by tuning the distance between the four rods. Despite such a simple design, topological edge and corner states [see Fig.~1(e)] can appear in the photonic band gap (PBG) between the first and the second bulk bands (denoted as PBG I). For isotropic PCs with $l_x=l_y\equiv l$, we find that the intra- and inter-unit-cell distances ($l_{intra}$ and $l_{inter}$) between the neighboring rods control the band topology. $l_{intra}=2l$ and $l_{inter}=a-2l$ where $a$ is the lattice constant. For $l_{intra}<l_{inter}$ the PBG I has trivial topology, whereas for $l_{intra}>l_{inter}$ the PBG I carries nontrivial topology as signified by the parity-inversion at the X point. The parities of the bands which are the eigenvalues of the inversion symmetry operator are defined in the same way as those in Ref. [41]. A transition appears at $l_{intra}=l_{inter}$ (i.e., $l=a/4$), where the PBG is closed by the band degeneracy on the Brillouin zone boundary lines (e.g., the MX line) [see Figs.~1(b)-1(d)].

The photonic band structure and the topological properties of the isotropic PC can be well approximated by tight-binding model as depicted in Fig.~1(a) of which the Hamiltonian is 
\begin{equation}         
\mathcal{H}(\bm{k})=
\left(                 
  \begin{array}{cccc}   
    0 & h_{12} & h_{13} & 0\\  
    h_{12}^\ast & 0 & 0 & h_{24}\\ 
    h_{13}^\ast & 0 & 0 & h_{34}\\ 
    0 & h_{24}^\ast & h_{34}^\ast & 0 \\
    \end{array}
    \right) .
    \label{eq:1}
\end{equation}
Here $h_{12}=t_a+t_b\mathrm{exp}(\mathrm{i}k_x)$, $h_{13}=t_a+t_b\mathrm{exp}(\mathrm{-i}k_y)$, $h_{24}=t_a+t_b\mathrm{exp}(-\mathrm{i}k_y)$, $h_{34}=t_a+t_b\mathrm{exp}(\mathrm{i}k_x)$, and ${\bm k}=(k_x, k_y)$.

The tight-binding parameters, $t_a, t_b$, reflect the intra- and inter-unit-cell couplings between the neighboring rods, respectively. It is known that the above tight-binding model has nontrivial topology as characterized by the 2D polarization 
${\bm P}$ where
\begin{equation}
P_i=-\frac{1}{(2\pi)^2}\int d^2\bm{k}\mathrm{Tr} [\hat{{\cal A}}_i] , \quad i=x,y
\end{equation}
where $(\hat{{\cal A}}_i)_{mn}(\textbf{k})=\mathrm{i}\bra{u_m(\textbf{k})}\partial_{k_i}\ket{u_n(\textbf{k})}$, where $m$, $n$ run over all occupied bands, $\ket{u_m(\textbf{k})}$ is the periodic Bloch function for the $m$th band. The 2D polarization is connected to the 2D Zak phase~\cite{2DSSH0} via $\theta_i=2\pi P_i$ for $i=x,y$. 2D Zak phase of the PC equals $(\pi, \pi)$ [i.e., ${\bm P}=(\frac{1}{2},\frac{1}{2})$] for  $t_a<t_b$ (i.e., $l_{intra}>l_{inter}$) and reveals that the PC is in topological nontrivial insulating phase, while it equals $(0, 0)$ for $t_a>t_b$ (i.e., $l_{intra}<l_{inter}$) where the PC is in the trivial insulating phase.  Beyond the isotropic tight-binding model, $P_x$ and $P_y$ can be different, resulting in nontrivial topological invariants ${\bm P}=(\frac{1}{2},0)$ or ${\bm P}=(0,\frac{1}{2})$, beside ${\bm P}=(\frac{1}{2},\frac{1}{2})$. We will show that only the last case can lead to both topological edge and corner states, whereas the first two cases can only support topological edge states but not corner states.


Consider a box-shaped boundary between the PC with ${\bm P}=(\frac{1}{2},\frac{1}{2})$ and the PC with ${\bm P}=(0,0)$ [see Fig.~1(e)]. There are boundaries along the $x$-  and $y$-  directions which support topological edge states due to the 2D Zak phase. For instance, the nontrivial Zak phase $\theta_x=\pi$ for each $k_y$ gives rise to the edge states on the boundaries along the $y$-  direction for the whole edge Brillouin zone $k_y\in [-\frac{\pi}{a},\frac{\pi}{a}]$. Similar bulk-edge correspondence works for the boundaries along the $x$- direction. We consider a ribbon structure where a strip of a nontrivial PC is sandwiched in between two PCs in trivial phase. Then the band structure for ribbon, namely the projected band structure contains the information of the lower dimensional edge states. We numerically simulated this structure and the dispersion of 1D edge state is clearly shown in Fig. 2. For the isotropic case, namely, the projected band structures for $k_x-$ and $k_y-$directions are the same due to the symmetry. Without loss of generality, we calculate the projected band structure along $k_x-$direction for the combined structure as shown in Fig. 1(e). The result shows that there are 1D edge states in the first band gap. We also label the frequency of one of the corner state as dashed line in Fig. 2 

\begin{figure}
\centering
\includegraphics[scale=0.35]{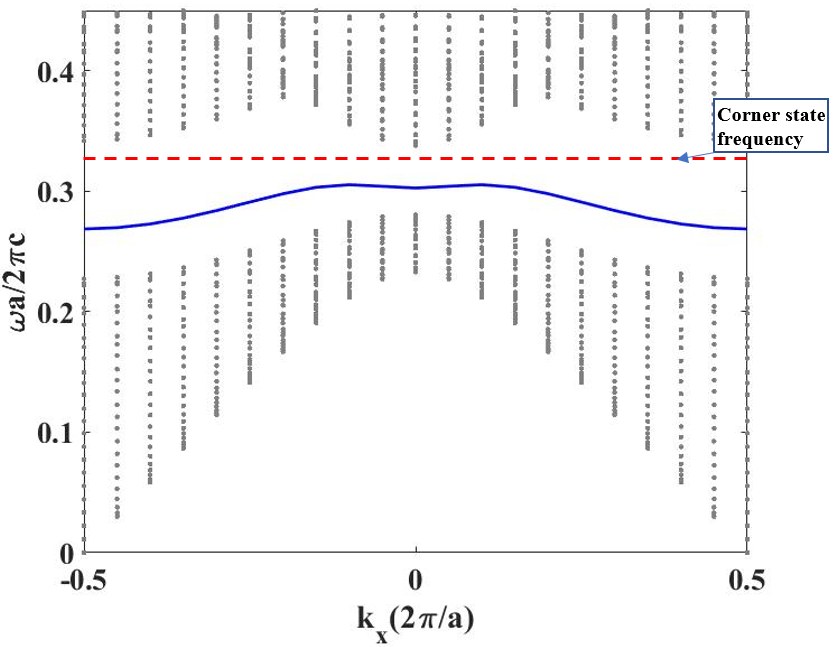}
\captionsetup{format=plain, justification=raggedright }
\caption{Projected band structure along $k_x-$direction for Fig. 1(e) (isotropic). There are 1D edge states appear in the first band gap which are topologically protected by bulk polarization  as depicted by solid blue line. The dashed red line labels the frequency of corner states.}
\label{fig:4}
\end{figure}

Remarkably, those 1D edge states themselves are similar to 1D photonic SSH model with nontrivial topology protected by $M_x$ ($M_y$) for the edges along the $x$- ($y$-) direction. The topology of the edge states is characterized by the edge polarizations, $p_y^{\nu_x}$ and $p_x^{\nu_y}$, for edges perpendicular to the $x$- and $y$- directions, respectively. 

The topological theory of polarization~\cite{HOTI1,HOTI4} connects the polarization of the edge states to the eigenvalues of the nested Wilson-loops (Mathematical details are given in the Supplemental Material). In simple terms, the nested Wilson-loops construct the Zak phases of the edge states by projecting the bulk Bloch states into a sector which is topologically equivalent to the edge states using the Wannier band basises, $\ket{w_{x,\alpha}({\bm k})}=\sum_{n\in {\cal N}_{PBG}}\nu^\alpha_n(k_y)\ket{u_n({\bm k})}$. Here $n\in{\cal N}_{PBG}$ stands for summation over all bands below the PBG, and $\nu^\alpha_n(k_y)$ is the $\alpha$-th Wilson-loop eigenvector for the Wilson-loop operator with $k_x$ looping from $-\pi/a$ to $\pi/a$ at fixed $k_y$. With the nested Bloch states, one can calculate the polarization for the edge states perpendicular to the $x$-  direction,
\begin{equation}
p_y^{\nu_x} = -\frac{1}{(2\pi)^2}\int d^2\bm{k}\mathrm{Tr} [\hat{{\cal B}}_y]  
\end{equation}
where $(\hat{{\cal B}}_y)_{\alpha,\beta}=\mathrm{i}\bra{w_{x,\alpha}({\bm k})}\partial_{k_y}\ket{w_{x,\beta}({\bm k})}$.
Remarkably, for the PBG considered in this work, there is only one photonic band below the PBG. Hence, there is only
one Wilson-loop eigenvalue and the eigenvector $\nu^1_1(k_y)\equiv 1$. Hence, the nested Wilson-loop becomes the same as the bulk Wilson-loop. However, the above equation is valid only when there is a physical edge states on the boundary perpendicular to the $x$ direction, i.e., only when $P_x=\frac{1}{2}$ (The edge polarization should vanish when $P_x=0$). Putting these two factors together, we find that $p_y^{\nu_x}=(2P_x)\times P_y=\frac{1}{2}$. Similarly, $p_x^{\nu_y}=(2P_y)\times P_x=\frac{1}{2}$. Numerical calculations of these topological indices are presented in the Supplemental Material~\cite{sup}.

The topological edge states along the $x$- and $y$- directions meet at the corner of the box-shaped combined structure shown in Figs.~1(e) and 3(b) where the outside PC has trivial topology, i.e., ${\mathbf P}=(0,0)$. Differing from Refs.~\cite{HOTI1,HOTIEXP1,HOTIEXP2,HOTIEXP3}, there is no quadrupole topological order in our system, since there is only a single band below the PBG. Therefore, the topological corner charge is determined by the edge polarizations as~\cite{HOTI4}
\begin{equation}
Q_c=p_x^{\nu_y} + p_y^{\nu_x}=4P_xP_y. \label{Qc}
\end{equation}
If the inner PC has ${\mathbf P}=(\frac{1}{2}, \frac{1}{2})$, then $Q_c=1$. This quantized corner charge gives rise to a single corner state in each of the four corners, as confirmed by the numerical simulation in Fig.~3(b). The detailed dielectric structure of the combined structure is given in Fig.~3(c) for one of the corner. The spectrum of the eigenmodes is already shown in Fig.~1(e) where four degenerate corner modes 
are found in the PBG. The electromagnetic fields of the corner modes are strongly localized at each of the four corners. These corner states are protected by the nontrivial topology, as characterized by $Q_c=1$, as well as the mirror symmetries $M_x$ and $M_y$ which quantize the polarizations. The bulk-edge-corner correspondence elucidated above shows the topological protection in a hierarchy of dimensions, which is a smoking-gun signature of HOTIs. The robustness of the corner states against perturbations is demonstrated by the simulations presented in the Supplemental Material~\cite{sup}.

\begin{figure}
\centering
\includegraphics[scale=0.16]{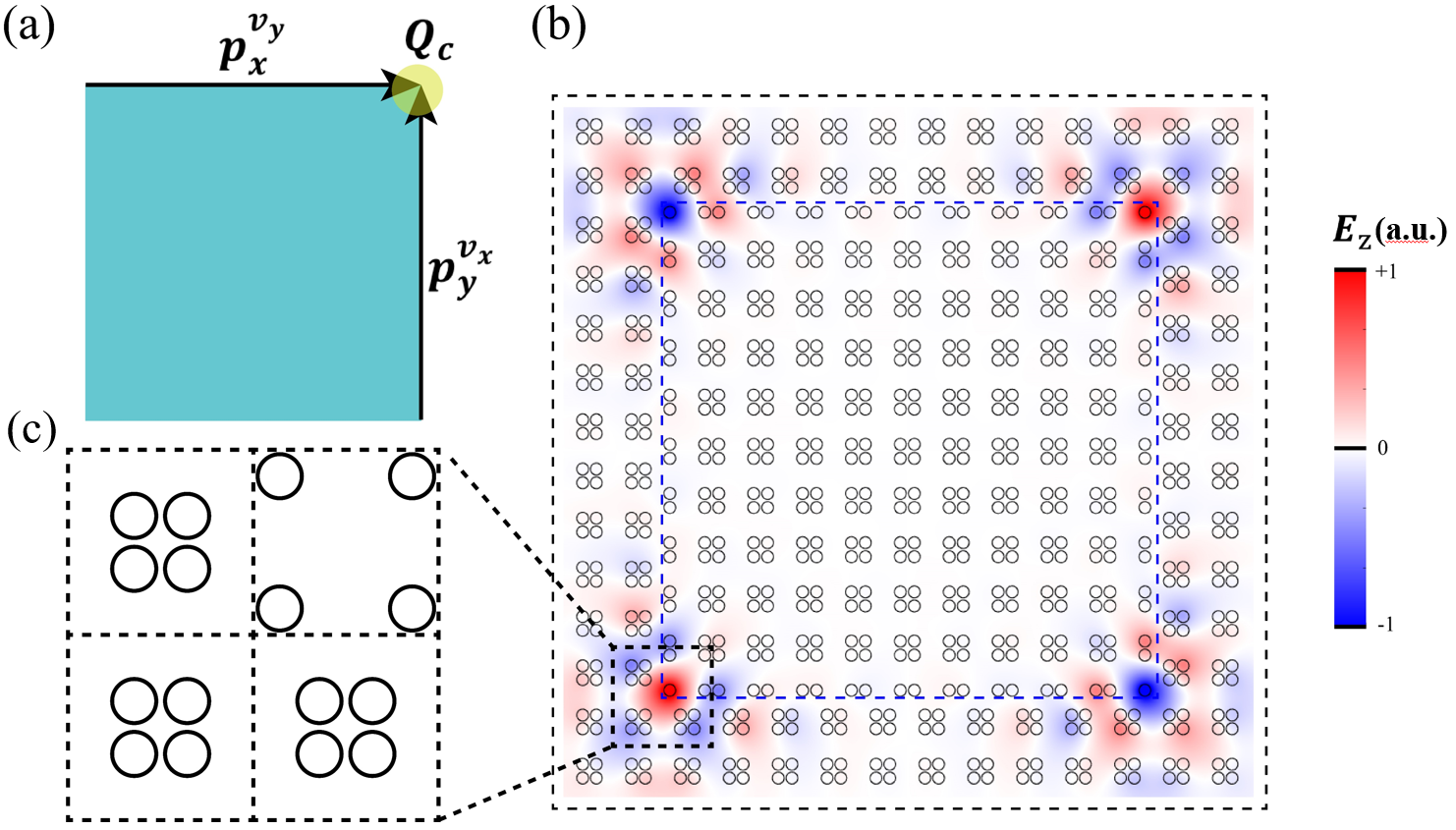}
\captionsetup{format=plain, justification=raggedright }
\caption{(a) Schematic of bulk-edge-corner correspondence. The nontrivial topology of the bulk (cyan region) leads to the emergence of the edge states, while the polarization of the edge (indicated by the black arrows) results in the formation of the topological corner state (indicated by the shallow yellow region). (b) There are in total four degenerate corner states at the four corners with frequency $66.0~\mathrm{THz}$. The electrical fields $E_z$ (a.u. stands for arbitrary units) are strongly localized at the corners (a superposition of the four degenerate corner states is shown here). The blue dashed line indicate the boundary between the two PCs. (c) The zoom-in structure of the corner, where the dashed black lines indicate the border of the unit-cells. The topological unit-cell has a much larger $l$ ($l=a/2.8$) compared to the three trivial unit-cells ($l=0.14a$).}
\label{fig:4}
\end{figure}

On the other hand, the emergence of the corner modes can be understood from the tight-binding model. Particularly, when one consider the extreme case where $t_a=0$ and $t_b\neq0$. In this case, the four corner sites of the tight-binding model become ``dangling atoms" which trap zero-energy corner modes. Since the band structure of this extreme case can be adiabatically connected to the $t_a\neq0$ case, they are in the same topological class and the corner states will always exist as long as $|t_a|<|t_b|$. In terms of the topological invariants, $Q_c=1$ corresponds to the ${\bm P}=(\frac{1}{2},\frac{1}{2})$ phase.

\begin{center}
\textbf{\uppercase\expandafter{\romannumeral3}. ANISOTROPIC PHOTONIC CRYSTALS AND TOPOLOGICAL PHASE DIAGRAM}
\end{center}

Next, we extend our discussions to anisotropic 2D PCs where
$l_x\neq l_y$ as shown in Fig.~4. The bulk polarizations are 
determined by the intra- and inter-unit-cell distance between the rods along the $x$- and $y$- directions, 
$l_{intra}^i$ and $l_{inter}^i$ with $i=x,y$. For instance, $l_{intra}^x>l_{inter}^x$ leads to nontrivial topological indices $P_x=\frac{1}{2}$. The edge polarizations are given by $p_x^{\nu_y}=p_y^{\nu_x}=2P_x P_y$.

For a box-shaped combined structure with the outer PC of ${\mathbf P}=(0,0)$, the edge states along the $y$- ($x$-) direction 
emerge when the inner PC has $P_x=\frac{1}{2}$ ($P_y=\frac{1}{2}$), as shown in Fig.~4. The electrical fields for the 
edge states in Fig.~4(a) indicate that the four dielectric rods at the corners have vanishing field intensity, emerging as
dangling atoms in the topological SSH model. Such unoccupied dielectric rods leave space for unpaired topological corner modes. In comparison, for inner PCs of ${\mathbf P}=(\frac{1}{2},0)$ or $(0, \frac{1}{2})$as shown in Fig.~4(b)-(c), the corner rods have finite field intensities and do not support corner modes. These observations are consistent with the corner charge given in Eq.~(\ref{Qc}). We also study the 1D edge states in the anisotropic cases. For the anisotropic case, namely  with x-direction in topological non-trivial phase and y-direction in topological trivial phase, the projected band structures are different for $k_x$- and $k_y$-directions. We calculate the projected band structures for Fig. 4(c) as a demonstration. The result is shown in Fig. 5(a) and Fig. 5(b) for the projected band structures along $k_x-$ and $k_y-$ direction respectively. The results of simulation match our theoretical predictions well.

\begin{figure}
\centering
\includegraphics[scale=0.24]{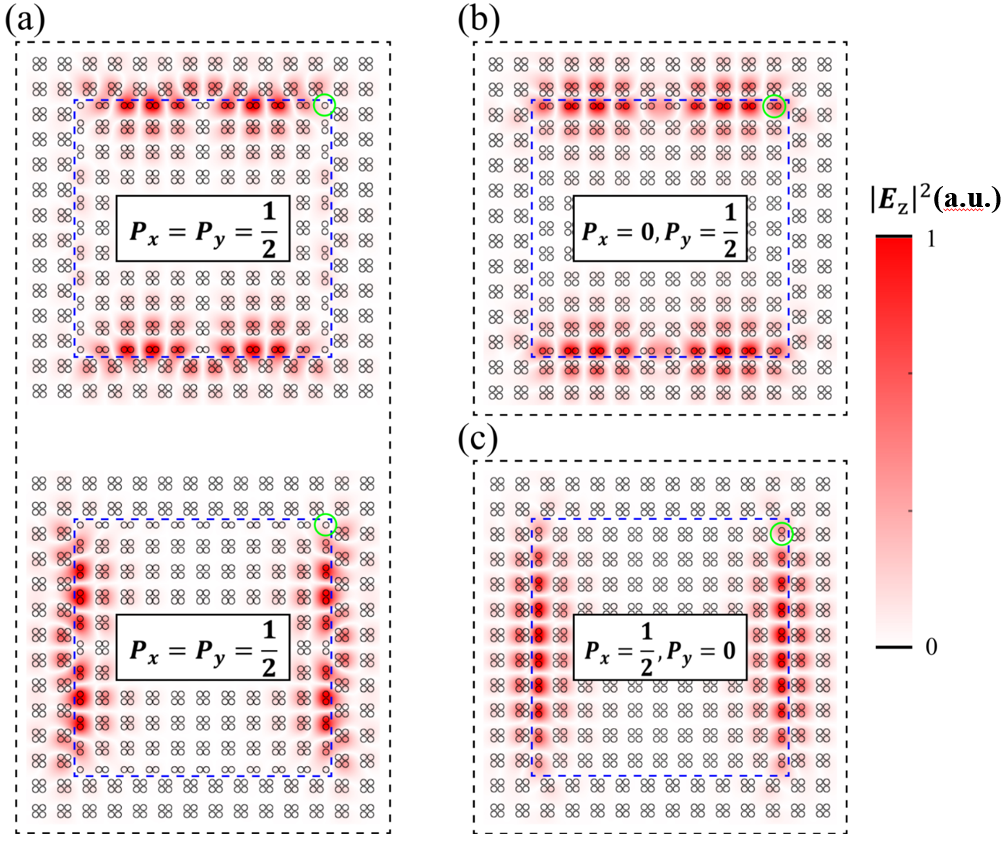}
\captionsetup{format=plain, justification=raggedright }
\caption{Anisotropic 2D photonic SSH model where $l_x\neq l_y$. The PCs have the same $a$, $r$ and $\epsilon$ as the isotropic 2D photonic SSH model but have deformed lattice structures characterized by $l_x$ and $l_y$. The eigenstates are solved by COMSOL. For outer PC, we set $l_x=l_y=0.14a$. For inner PCs, we set (a) $l_x=0.37a$ and $l_y=0.35a$. Non-trivial phase in both $x$- and $y$- directions. Two sets of DESs with different frequencies. For the DESs in the upper (lower) figure, the frequency is $60.2\mathrm{THz}$ ($61.1\mathrm{THz}$). (b) $l_x=0.15a$ and $l_y=0.35a$. $x$-direction in topologically trivial phase and $y$-direction in topologically non-trivial phase. There are only 1D DESs in the upper and lower sides. The frequency is $64.2\mathrm{THz}$. (c) $l_x=0.35a$ and $l_y=0.15a$. $x$-direction in topologically non-trivial phase and $y$-direction in topologically trivial phase. There are only 1D DESs in the left and right sides. The frequency of the state in this figure is $65.2\mathrm{THz}$. The solid green circles emphasize the differences of the corner structures and field strengths in (a)-(c). The dashed blue lines label the boundaries of two PCs.}
\label{fig:6}
\end{figure} 

\begin{figure}
\centering
\includegraphics[scale=0.26]{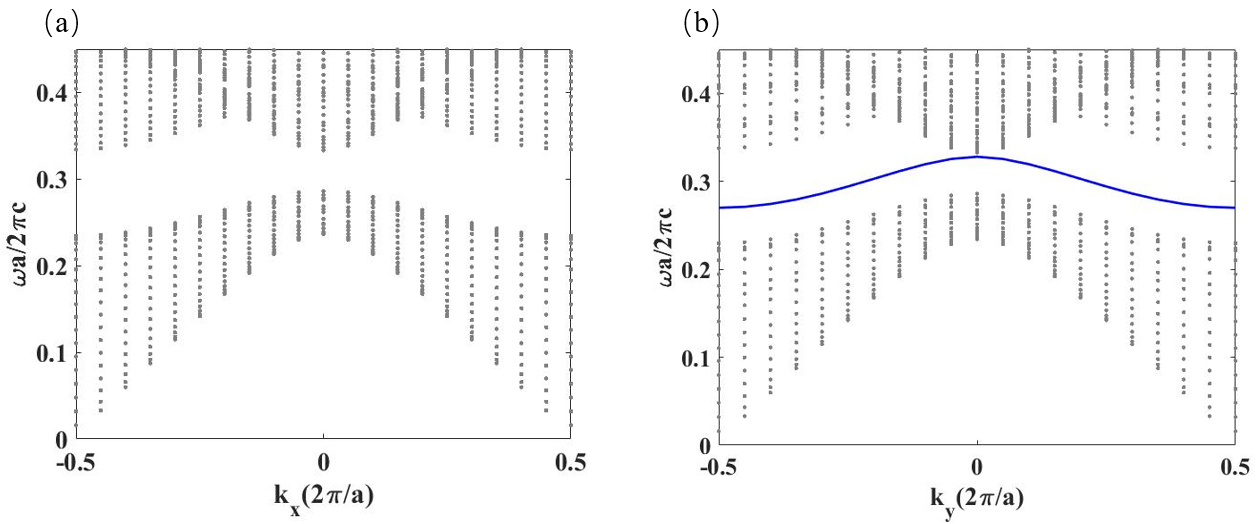}
\captionsetup{format=plain, justification=raggedright }
\caption{Projected band structures for Fig. 4(c) along $k_x-$ and $k_y-$direction as shown in (a) and (b) respectively. (a) There are no 1D edge states along $k_x-$direction. (b) There are 1D edge states in the band gap along $k_y-$direction as depicted by solid blue line.}
\label{fig:6}
\end{figure} 

According to the above discussions, the whole phase diagram of the 2D PCs can be classified into four different phases, as depicted in Fig.~6: the topologically trivial phase with $l_{intra}^i<l_{inter}^i$ for both $i=x$ and $y$ 
of which ${\bm P}=(0,0)$ and $Q_c=0$; the two phases with nontrivial bulk topology but zero corner charge, i.e., ${\bm P}=(\frac{1}{2},0)$ or $(0, \frac{1}{2})$, and $Q_c=0$ where $l_{intra}^x<l_{inter}^x$ or $l_{intra}^y<l_{inter}^y$; the phase with both nontrivial bulk topology and corner charge, i.e., ${\bm P}=(\frac{1}{2},\frac{1}{2})$ and $Q_c=1$ where $l_{intra}^i>l_{inter}^i$ for both $i=x$ and $y$. Therefore, by simply changing the relative distances of nearest rods in anisotropic 2D photonic SSH model, we can achieve various topological phases with different 1D DESs as well as corner states. This can be potentially used to design novel optical topological switch for photonic integrated chips in future.

\begin{figure}
\centering
\includegraphics[scale=0.21]{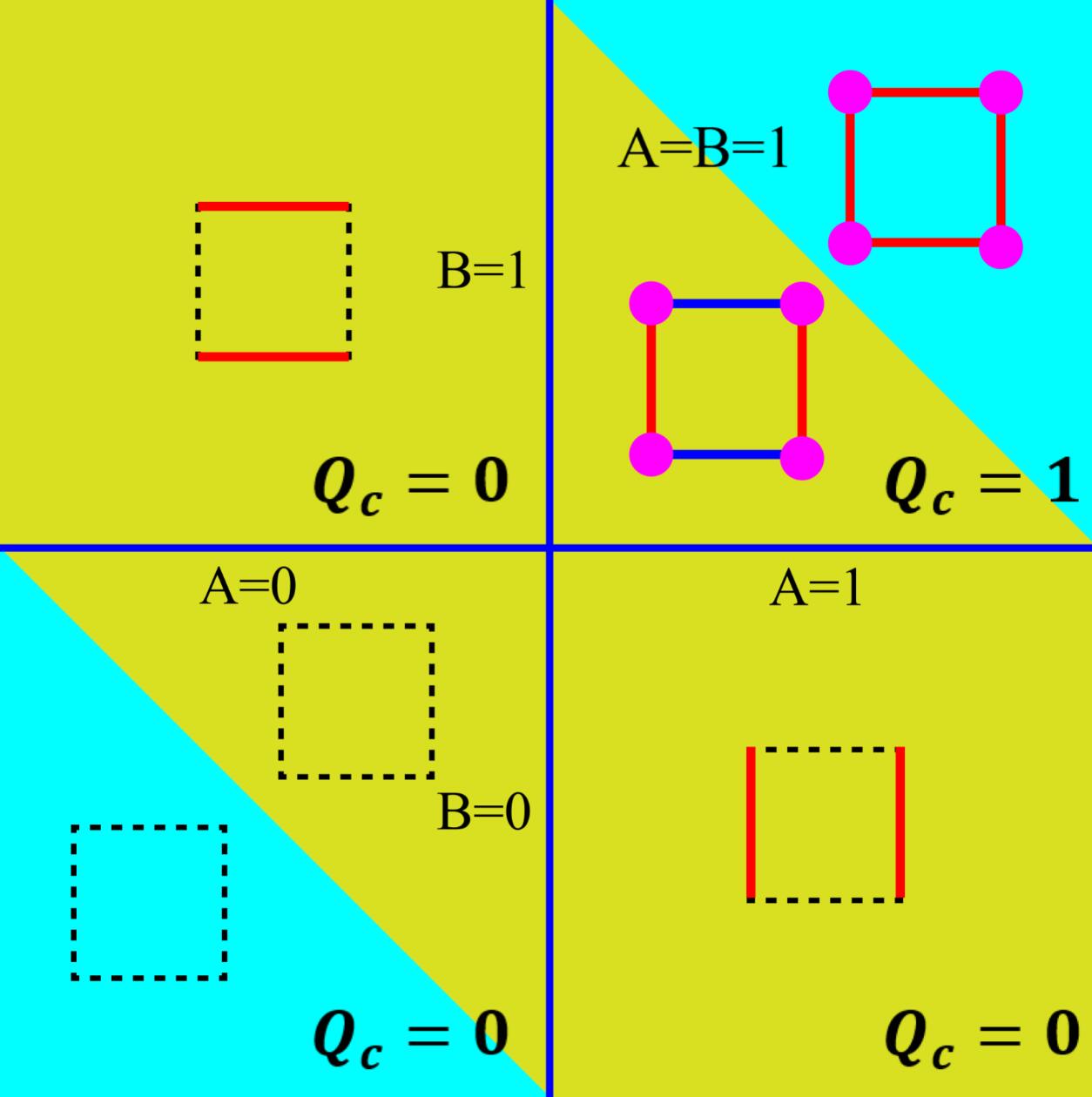}
\captionsetup{format=plain, justification=raggedright }
\caption{Classification of all kinds of  generalized 2D photonic SSH model with different configurations in $x$-and $y$-directions. $\mathrm{A (B)} =1$ means 2D photonic SSH model is in topologically nontrivial phase along $x$-($y$-) directions. $\mathrm{A (B)} =0$ means 2D photonic SSH model is in topologically trivial phase along $x$-($y$-) direction. The blue area represents isotropic cases and yellow area represents anisotropic cases. The dashed lines represent there are no 1D edge states and solid lines represent the existence of 1D edge states. The solid circles means there are corner states. The edge (or corner) states of the same color have the same frequency, whereas the states with different colors have different frequencies. }
\label{fig:7}
\end{figure} 

 \begin{center}
\textbf{\uppercase\expandafter{\romannumeral4}. CONCLUSIONS AND DISCUSSIONS}
\end{center}

We propose a simple realization of the second-order topological insulator in all-dielectric photonic crystals (PC) with corner states. The exotic corner states can be regarded as the 0D boundary states of the 1D edge of the 2D PC which is topologically protected by mirror symmetries. Besides, we study the anisotropic 2D photonic SSH model and find that the 1D DESs arise due to the 1D structures existing in 2D photonic SSH model. By adjusting the distances between the nearby rods in the $x$-and $y$-directions, the emergence of the edge and corner states can be controlled straightforwardly. 

These topologically protected edge and corner states may be valuable for robust waveguides, optical couplers, and optical topological circuit switches. Our PC can be experimentally realized in surface plasmon polariton (SPP) slab and in microwave frequencies. In both cases, the structure is similar to our model but has a finite size in the z-direction. The relevant parameters need to be modified from our theoretical model but the topological properties are the same.

If our theory is generalized to 3D photonic SSH model, there can be second-order and third-order topological insulating phases where topological hinge states and corner states can emerge respectively. 3D second-order topological semimetals may be achieved by stacking the 2D photonic SSH model along the $z$-direction. The 3D second-order topological semimetals have rich bulk and boundary properties which are yet to be explored~\cite{HOTSM}.

{\sl Note added}. A recent paper showing 0D topological bound state in a 2D PC due to the presence of dislocation (a topological defect) has appeared~\cite{dislocation}, demonstrating another mechanism for lower-dimensional topological light-trapping.

{\sl Acknowledgments}.
B.Y.X, H.F.W, X.Y.Z, M.H.L and Y.F.C are supported by the National Key R$\&$D Program of China (Grant No. 2017YFA0303700, 2018YFA0306200) and the National Nature Science Foundation of China (Grant No. 51721001) as well as the Academic Program Development of Jiangsu Higher Education (PAPD). H.X.W and J.H.J thank supports from the National Natural Science Foundation of China (No. 11675116).


\begin{thebibliography}{References}

\bibitem{TI1} M. Z. Hasan and C. L. Kane, Rev. Mod. Phys. \textbf{82}, 3045 (2010).

\bibitem{TI2} X.-L. Qi and S.-C. Zhang, Rev. Mod. Phys. \textbf{83}, 1057 (2011).

\bibitem{PTI1} F. D. M. Haldane and S. Raghu, Phys. Rev. Lett.  {\bf 100}, 013904 (2008). 

\bibitem{PTI2} Z. Wang, Y. Chong, J. D. Joannopoulos, and M. Solja\v{c}i\'{c}, Nature  {\bf 461}, 772-775 (2009).

\bibitem{PTI3} M. Hafezi, E. A. Demler, M. D. Lukin, and J. M. Taylor, Nat. Phys. {\bf 7}, 907-912 (2011). 

\bibitem{PTI4} Y. Poo, R. X. Wu, Z. Lin, Y. Yang, and C. T. Chan, Phys. Rev. Lett. {\bf 106}, 093903 (2011). 

\bibitem{PTI5} A. B. Khanikaev {\sl et al.} Nat. Mater.  {\bf 12}, 233-239 (2013). 

\bibitem{PTI6} M. Hafezi, S. Mittal, J. Fan, A. Migdall, and J. M. Taylor, Nat. Photon.  {\bf 7}, 1001-1005 (2013). 

\bibitem{PTI7} M. C. Rechtsman  {\sl et al.}  Nature  {\bf 496}, 196-200 (2013). 

\bibitem{PTI8} L. Lu,  J. D. Joannopoulos, and M. Solja\v{c}i\'{c}, Nat. Photon. {\bf 8}, 821-829 (2014).

\bibitem{PTI9} W.-J. Chen {\sl et al.}  Nat. Commun. {\bf 5}, 6782 (2014).

\bibitem{PTI10} S. A. Skirlo, L. Lu, and M. Solja\v{c}i\'{c}, Phys. Rev. Lett.  {\bf 113}, 113904 (2014).

\bibitem{PTI11} W. Gao, M. Lawrence, B. Yang, F. Liu, F. Fang, B. Beri, J. Li, S. Zhang, Phys. Rev. Lett. {\bf 114}, 037402 (2015).

\bibitem{PTI12}L.-H. Wu and X. Hu, Phys. Rev. Lett. \textbf{114}, 223901 (2015).

\bibitem{PTI13} T. Ma, A. B. Khanikaev, S. H. Mousavi, and G. Shvets, Phys. Rev. Lett. {\bf 114}, 127401 (2015). 

\bibitem{PTI14} C. He {\sl et al.}  Proc. Natl. Acad. Sci. USA  {\bf 113}, 4924-4928 (2016). 

\bibitem{PTI15} D. Leykam, M. C. Rechtsman, and Y. D. Chong, Phys. Rev. Lett. {\bf 117}, 013902 (2016). 

\bibitem{PTI16} F. Gao {\sl et al.}  Nat. Commun.  {\bf 7}, 11619 (2016). 

\bibitem{PTI17}L. Xu, H.-X. Wang, Y.-D. Xu, H.-Y. Chen, and J.-H. Jiang, Opt. Express \textbf{24}, 18059 (2016).

\bibitem{PTI18} X.-C. Sun, C. He, X.-P. Liu, M.-H. Lu, S.-N. Zhu, and Y.-F. Chen, Prog. Quant. Electron. \textbf{55}, 52 (2017).

\bibitem{PTI19} Xuan Zhu, Hai-Xiao Wang, Changqing Xu, Yun Lai, Jian-Hua Jiang, and Sajeev John, Phys. Rev. B {\bf 97}, 085148 (2018).

\bibitem{PTI20} Y. Yang, Y. F. Xu, T. Xu, H.X. Wang, J. H. Jiang, X. Hu, Z. H. Hang, Phys. Rev. Lett. {\bf 120}, 217401 (2018).

\bibitem{PTSM1} L. Lu, L. Fu, J. D. Joannopoulos, and M. Solja\v{c}i\'{c}, Nat. Photon. {\bf 7}, 294-299 (2013).

\bibitem{PTSM2} L. Lu, Z. Wang, D. Ye, L. Ran, L. Fu, J. D. Joannopoulos, and M. Solja\v{c}i\'{c}, Science \textbf{349}, 622 (2015).

\bibitem{PTSM3} H.-X. Wang, L. Xu, H.-Y. Chen, and J.-H. Jiang, Phys. Rev. B {\bf 93}, 235155 (2016).

\bibitem{PTSM4} Meng Xiao, Qian Lin, and Shanhui Fan
Phys. Rev. Lett. {\bf 117}, 057401 (2016).

\bibitem{PTSM5} H.-X. Wang, Y. Chen, Z. H. Hang, H.-Y. Kee, and J.-H. Jiang, npj Quantum Materials {\bf 2}, 54 (2017).

\bibitem{PTSM6} B. Yang {\sl et al.} Science {\bf 359}, 1013-1016 (2018).

\bibitem{HOTI1} W. A. Benalcazar, B. A. Bernevig, and T. L. Hughes,
Science {\bf 357}, 61-66 (2017).

\bibitem{HOTI2} Z. Song, Z. Fang, and C. Fang, Phys. Rev. Lett. \textbf{119}, 246402
(2017).

\bibitem{HOTI3} J. Langbehn, Y. Peng, L. Trifunovic, F. von Oppen, and
P.W. Brouwer, Phys. Rev. Lett. \textbf{119}, 246401 (2017).

\bibitem{sup} See Supplemental Material.

\bibitem{HOTI4} W. A. Benalcazar, B. A. Bernevig, and T. L. Hughes, Phys.
Rev. B \textbf{96}, 245115 (2017).

\bibitem{HOTI5} F. Schindler, A. M. Cook, M. G. Vergniory, Z. Wang, S. S. P.
Parkin, B. A. Bernevig, and T. Neupert, arXiv:1708.03636.

\bibitem{HOTI6} M. Ezawa. Phys. Rev. Lett. \textbf{120} 026801 (2018).


\bibitem{HOTIEXP1} M. Serra-Garcia, V. Peri, R. Susstrunk, O. R. Bilal, T. Larsen, L. G. Villanueva, S. D. Huber, Nature \textbf{555}, 342 (2018).

\bibitem{HOTIEXP2} C. W. Peterson, W. A. Benalcazar, T. L. Hughes, and G. Bahl
Nature {\bf 555},  346-350 (2018).

\bibitem{HOTIEXP3}S. Imhof, C. Stefan, F. Bayer, J. Brehm, L. Molenkamp, T. Kiessling, F. Schindler, C. H. Lee, M. Greiter, T. Neupert, R. Thomale, arXiv:1708.03647 (2017).

\bibitem{HOTIEXP4} J. Noh {\sl et al.} arXiv:1611.02373 (2016). 

\bibitem{SSH} W. P. Su, J. R. Schrieffer, and A. J. Heeger, Phys. Rev. Lett.  {\bf 42}, 1698-1701 (1979). 


\bibitem{2DSSH0} F. Liu, and K. Wakabayashi, Phys. Rev. Lett. \textbf{118}, 076803 (2017).

\bibitem{2DSSH1} F. Liu, H. Y. Deng, and K. Wakabayashi. Phys. Rev. B \textbf{97}, 035442 (2018).

\bibitem{HOTSM} M. Lin and T. L. Hughes, arXiv:1708.08457.

\bibitem{dislocation} F.-F. Li et al.  arXiv:1802.01811 (2018).

\end{thebibliography}
\end{document}